\documentstyle[twocolumn,prl,aps]{revtex}
\tighten
\begin{document}

\draft
\title{A Universal Nucleation Mechanism for Solvent Cast Polymer Film Rupture}
\author{D. Podzimek, A. Saier, R. Seemann, K. Jacobs, and
S. Herminghaus}
\address{Universit\"at Ulm, Abteilung Angewandte Physik, D-89069 Ulm, Germany}
\maketitle
\begin{abstract}
It is shown that the intrinsic stress in solvent cast polymer
coatings plays a key role in the nucleation of holes in the film.
Nucleation is important because it is meanwhile clear that
heterogeneous nucleation is the only relevant rupture mechanism
for the technologically relevant thickness regime well above 100
nm. The most striking feature is that in contrast to what has been
widely believed, the number density of holes scales not
algebraically, but {\it exponentially} with the film thickness.
\end{abstract}
\vspace{0.5cm} If one deposits a thin liquid film on a solid
substrate, such as a freshly prepared laquer coating, one usually
wishes it to stay homogeneously in place. However, if the
equilibrium contact angle of the film material is finite, the film
can gain free energy by beading off the substrate and forming an
array of individual droplets. In order to avoid this dewetting
process, a thorough understanding of the basic mechanisms is
necessary.

For films thinner than about 100 nanometer, the initial formation
of dry spots may proceed via three distinctly different
mechanisms. Dry patches may nucleate at impurities or
inhomogeneities (heterogeneous nucleation), or by thermally
excited indentations in the free film surface (thermal
nucleation). Furthermore, capillary waves on the film may be
unstable and grow in amplitude until their troughs reach the
substrate, where they form dry spots (spinodal dewetting). The
interplay of these dewetting modes has only recently come close to
a complete understanding \cite{See1,See2}.

For the technologically much more relevant thickness regime well
above 100 nm, it is meanwhile clear that heterogeneous nucleation
is the only relevant mechanism for the overwhelming majority of
systems. In this scenario, a random array of circular holes form
immediately, one at each nucleation site \cite{Jacobs98}. They
grow in time until all of the film has been removed. If the
nucleation centers could be completely removed from the film, the
latter would be metastable, and usually last sufficiently long for
most applications. It is therefore of particular importance to
investigate the very nature of the nucleation centers. Since the
most relevant material class in this context are polymers
deposited from solution, we used solvent cast polystyrene films on
silicon wafers as a model system, and we restrict our discussion
to this frequently encountered type of film throughout this paper.

The most obvious project is to try to reduce the number of
nucleating holes by cleaning up the film material and the
preparation conditions. However, the amazing gap observed between
enormous cleaning efforts on the one hand, and the poor success on
the other, suggests that the physics of hole nucleation in polymer
films may be deeper than a mere effect of `dirt'. We have thus
investigated the possible interplay between the nucleation of
holes in the film, and some possible strain in the material
induced by the preparation procedure.

In fact, it is well known that solvent cast polymer films are
considerably strained by means of a quite fundamental mechanism:
as the solvent evaporates and the film becomes thinner, the glass
transition temperature, $T_g$, of the solution gradually increases
from well below room temperature to the glass transition
temperature of the pure polymer, which is about $T_g =
100^{\circ}$C for polystyrene. As long as $T_g$ is below room
temperature, the polymer molecules are mobile and retain their
equilibrium configuration, the gyration ellipsoids of the
individual molecules averaging to a sphere. However, as $T_g$
increases above room temperature, equilibration of the molecules
is hindered, such that the (averaged) gyration ellipsoid is
flattened as the film thins further. The strain generated in this
way is thus determined by the solvent content of the film when
$T_g$ equals room temperature, and is largely independent of other
parameters, such as the rate of evaporation, or the film
thickness. For instance, the stress in a polystyrene film cast
from toluene at about 20$^{\circ}$ C is, quite universally, 14 MPa
\cite{Croll79} .

In order to investigate the dewetting behavior of polystyrene
films with different strain but otherwise identical properties, we
had to find a way to vary the strain in a controlled manner.
Polystyrene films were thus spin cast from toluene solution onto
mica substrates. Since polystyrene wets mica completely, the films
could be annealed on the mica substrate without any formation of
holes, and the annealing step should change the amount of stress
in the film. Afterwards, the annealed films were floated onto
millipore water and transferred to hydrophobized silicon wafers
\cite{footnote1} and heated above $T_g$, such that dewetting could
commence. The formation of holes was observed with an optical
reflection microscope.

In Fig.~\ref{annealing}, we show the number density of holes
forming in a 60 nm thick film of 52~kg/mol polystyrene
\cite{footnote2}, as a function of the annealing time. Each data
point corresponds to several prepared samples, the error bars
represent standard deviations. Annealing was done at
140$^{\circ}$C, which is 40 degrees above the bulk glass
transition $T_g$ temperature. As one can clearly see, the number
of holes decreases significantly with increasing annealing time.
It is important to mention, though, that a reduction of hole
density cannot be achieved by reducing the residual solvent
content of the film, for instance by storing the samples under
vacuum or by annealing them to a temperature below $T_g$. Only
annealing above $T_g$ will result in a reduction. This shows that
the mechanism of hole nucleation must be intimately connected to a
material property of the film which is reduced upon annealing,
such as the intrinsic strain. The solid line is an exponential fit
to the data, indicating a decay time of $\tau_0 = (240 \pm 20)$ s
Within our experimental scattering, the data are also consistent
with a stretched exponential, $exp(-t^{\beta})$, with $\beta$
between $0.8$ and $1.1$. However, since no significant deviation
from a pure exponential is observed, the decay will henceforth be
characterised by a single decay time, assuming $\beta = 1$.

It is clear from Fig.~\ref{annealing} that there is a residual
subset of nucleation sites which is not removed by annealing. We
will refer to the number density of these sites as $n_0$. However,
a significant reduction of the number density of holes, as
displayed in the figure, was invariably observed, and deserves
thorough investigation. We thus repeated the procedure with films
of 390 kg/mole polystyrene, and found a decay time of $\tau_0 =
(510 \pm 200)$ s. The ratio of the two decay times is close to the
ratio of the self diffusion constants determined by NMR for these
molecular weights \cite{Bac83}. This corroborates that microscopic
changes in molecular configuration are probably responsible for
the drastic changes in the number density of nucleation sites.

Let us now discuss the dependence of the number density $n_s$ of
nucleation sites which are deactivated or removed by annealing. It
corresponds to the difference of the initial number density of
holes and the number density found after long annealing times. It
is particularly rewarding to plot this quantity as a function of
film thickness, $h$. This is done in Fig.~\ref{thickness}, which
shows the data along with a simple exponential fit. Good agreement
is found, over a range of five orders of magnitude in number
density. This suggests a fundamental law behind the nucleation of
holes, and calls for explanation by a theoretical model.

What may be the microscopic nature of defects causing a local
reduction in strain? A straightforward model is to assume these to
be defects in the entanglement of the polymer chains. Regions of
(accidentially) reduced entanglement will always be present, as a
matter of mere statistics, and the material will be `weaker'
there. But how can an object as gentle as an entanglement defect
induce an indentation going all the way through the film and
finally reaching the substrate, forming a dry patch? It very
probably won't! Instead, quite a number of such defects must be
present within the size of a typical indentation, in order to
create an indentation deep enough to reach the substrate and
initiate dewetting. If the density of these defects is constant
within the polymer, as it is to be assumed, the probability of a
certain number of defects being accumulated within the size of an
indentation, which is the condition of forming a hole, decays
exponentially with the film thickness. This is a quite robust
result, and within our model explains the exponential law
displayed in Fig.~\ref{thickness}.

As a final test of the ideas presented here, let us discuss what
we would expect for the dependence of the density of holes as a
function of molecular weight. As the latter is reduced below the
limit of entanglement, the defects must vanish. Consequently, we
expect, roughly and qualitatively, that the number of holes
increases with increasing molecular weight. In
Fig.~\ref{molweight}, we have plotted the density of holes in
films of 60 nm thickness for variable molecular weight. The
increasing tendency is clearly corroborated. An investigation of
the fact that the data are consistent with a linear relationship,
however, goes beyond the scope of this paper, and must be left to
further studies.

In conclusion, we have obtained a clear insight into the nature of
one class of nucleated holes. We present a model where nuclei,
sites of high stress, are responsible for one class of holes in
the dewetting film. These sites of high stress are intrinsic to a
polymer melt that is prepared out of a solvent solution, e.g. by
spin coating, dip coating or spraying. By annealing the melt above
glass transition temperature on top of a wettable substrate, the
melt is able to release the stress and the polymer chains can
attain to their radius of gyration typical in the melt. After the
transfer of the film onto a non-wettable substrate, number density
of holes present after annealing was substantially reduced. The
model is corroborated by the results of experiments with
long-chained polymer molecules. Here, the number density of
nucleation sites which are `deactivated' by annealing increases
with increasing polymer chain length, suggesting a direct
proportionaliy of number of entanglements and stress. Moreover,
our results suggest a way of stabilizing coatings for practical
use by reducing the stress inside a film by means of annealing or
by using short-chained polymers.

\acknowledgements This work was funded by the German Research
Society in the framework of the Priority Program `Wetting and
Structure Formation at Interfaces'. We acknowledge generous
support of Si wafers by Wacker Chemitronics, Burghausen, Germany.

\begin{figure}
\caption{ Density of holes in 60 nm thick PS(51.5k) film as
function of the annealing time. The solid line represents an
exponential fit to the data. } \label{annealing}

\caption{Density of nucleation sites $n_s$ which are deactivated
by annealing as function of film thickness. The solid line
represents an exponential fit to the data. } \label{thickness}

\caption{ Density of nucleation sites $n_s$ which are deactivated
by annealing as function of molecular weight. The solid line
represents a linear fit to the data.} \label{molweight}
\end{figure}


\begin{references}
\bibitem{footnote1}The hydrophobized silicon wafers used are covered by a
self assembly layer of octadecyl-trichlorosilane following
standard procedures \cite{Wassermann89,Brzoska94}.
\bibitem{footnote2}Polystyrene is purchased from
Polymer Standard Mainz, German with a molecular weight between 18
and 780 kg/mol, and Mw/Mn between 1.02 and 1.05.
\end{references}
\end{document}